# Exchange bias training effect in IrMn-layer/ferromagnetic-ribbon heterostructures probed with magnetoimpedance

Mohammadreza Hajiali[*,†], Mohammad Javad Kamali Ashtiani[†] , Loghman Jamilpanah and Majid Mohseni[*]

*Faculty of Physics, Shahid Beheshti University, Evin, 19839 Tehran, Iran*

**Abstract**

The exchange-bias training effect in IrMn-layer/ferromagnetic-ribbon heterostructure is studied by performing magnetoimpedance (MI) measurements. Asymmetric, hysteretic and single peak behavior of the MI response and a shift in the MI peak to zero fields accompanied by 52% increase in MI has been detected as the signature of the exchange bias (EB) and training effect (TE), respectively. Also, during the consecutive filed sweep of MI response, both EB field and the degree of asymmetry of MI decrease that is another reason for existence of TE in our sample. The analysis of the magneto-optical Kerr effect (MOKE) and MI behavior establish that Hoffmann's model (Phys. Rev. Lett. 93, 097203 (2004)) is a good description of our experimental data, due to the existence of a strong single cycle TE in our IrMn-layer/ferromagnetic-ribbon system.

---

[*] Corresponding author's email address: mrh.hajiali67@gmail.com, m-mohseni@sbu.ac.ir

[†] These authors contributed equally to this work.

## Introduction

Direct exchange coupling at the interface of a ferromagnet (FM) and an antiferromagnet (AFM) can lock magnetic moments of the FM in a well-defined direction, making a unidirectional magnetic anisotropy at the FM/AFM interface [1–3]. This effect is also known as exchange bias (EB) which is mainly characterized by finding a shift in the hysteresis loop ($H_{EB}$) along the magnetic field axis and is usually accompanied by an enhancement of coercivity ($H_c$). This interfacial coupling has been studied intensively in the past two decades due to its unique features and applications in spintronic devices [4] and sensors such as giant magnetoresistance (GMR) read heads [5], magnetic random access memory (MRAM) and spin transfer torque MRAMs [6]. Conventionally, the direction and strength of the EB is determined with an in-plane field applied during growth or field cooling below the ordering temperature of the AFM [7]. Beside the hysteresis loop shift and $H_c$ enhancement, the EB phenomenon also exhibits asymmetry in the magnetization reversal process and training effect [8–12].

Training effect (TE) is one of the important phenomenon linked to the EB effect in which interfacial spin disorder and frustration occur particularly at low temperatures [13,14]. The TE refers to the gradual and monotonous degradation of both $H_{EB}$ and $H_c$ to those of equilibrium values during consecutive hysteresis loop measurements after field growth or field cooling. It is generally accepted that, TE arises due to irreversible changes in the magnetic microstructure of the AFM, as its spin structure rearranges with each magnetization reversal of the FM layer. Based on the behavior of the first and second consecutive hysteresis loop measurement, two types of TE can be seen as nominated thermal and athermal effects [15]. The thermal training is related to thermally activated depinning of the uncompensated AFM spins and it usually corresponds to a small change in both $H_{EB}$ and $H_c$ during each hysteresis cycle [16]. In contrast, athermal TE is virtually temperature independent and characterized by an abrupt decrease of $H_c$ and $H_{EB}$ in the first training, which is reported particularly in EB systems containing high symmetry and biaxial AFM anisotropy layers such as CoO [10] and IrMn [17].

Generally, the TE is largely depends on the properties of the AFM layer such as the microstructure and defects [18], field cooling procedure [19], strength of the interfacial exchange interaction [20] and temperature [21]. Experimentally, various techniques have been used to investigate EB and origin of TE such as polarized neutron reflectivity (PNR) [22], X-ray magnetic circular and Linear dichroism (XMCD, XMLD) [23,24], anisotropic magnetoresistance (AMR) [12,25], conventional magnetometry [26] and very recently Lorentz transmission electron microscopy (L-TEM) [27]. Indeed, simple experiments for the observation and comprehensive study the evolution of the EB and the TE in FM/AFM heterostructures is missing.

In this work, we use magnetoimpedance (MI) effect as a probe of the EB and the TE in in FM/AFM heterostructures that is made of an amorphous FM $Co_{68.15}Fe_{4.35}Si_{12.5}B_{15}$ ribbon and a thin layer of polycrystalline $Ir_{20}Mn_{80}$ (IrMn). The MI effect is a classical electrodynamic phenomenon in conducting FM with high transverse magnetic permeability ($\mu_t$) as the electrical impedance changes against external dc magnetic field [28–37]. The MI is correlated with the skin depth ($\delta = (\rho/\pi\mu_t f)^{1/2}$), of the high frequency $f$ current and $\mu_t$ of the conducting FM with electric resistivity $\rho$. Accordingly, $\mu_t$ changes by applying a dc external field and this results in a new current skin depth, thus varying the MI response.

Here, we present effect of sputter deposited IrMn layer on the MI response. From far viewpoints, as the FM ribbon is a thick structure with thickness of 20 µm, it is rather unexpected to observe the EB effect and hence the TE. However, the MI was shown in many literatures [29,30,32,34] to be a surface sensitive measurement technique with confined of the most ac current at 100 nm at the skin depth. Due to their high magnetic permeability, MI effect in such ribbons is very sensitive against tiny changes at the surface. We showed that impedance spectroscopy can be used for detection of the spin-orbit torque resulting from the spin Hall effect in a Pt-layer/magnetic-ribbon heterostructures [30]. Moreover, older studies have shown asymmetric MI effect made by intense surface oxidation of such ribbons [38]. To our knowledge, there is no report on the EB and TE probed via the MI effect. The result show that the IrMn layer creates a bias field in the ferromagnetic-ribbon that shifts the hysteresis loop and the MI response and also creates asymmetric and hysteretic behavior in the MI response. During the training procedure in field sweep MI measurement, both EB field and the degree of asymmetry of MI decrease. Our results can be used for development of simple methods for study of fundamental and experimental exchange bias training effect phenomena.

**Experimental**

Amorphous Co-based FM ribbons (1 mm width, 40 mm length and ~20 µm thickness) was prepared by a conventional melt-spinning technique. Before deposition of the IrMn layer, about 40 nm of the ribbons surface was etched via Ar plasma to have a clean and oxygen free surface. Immediately after that, the IrMn thin layer with thickness of 20 nm was deposited on the soft surface (shiny side) of those ribbons in the presence of the Ar with a pressure of 5 mTorr, base pressure better than $5\times10^{-6}$ Torr and growth rate of 3 nm/minute. In order to induce the EB, the as-deposited sample was subsequently annealed at 280 °C for 1 h under vacuum conditions ($4\times10^{-3}$ Torr) in the presence of in-plane magnetic field and then cooled down to room temperature. The magnetic field during the annealing and cooling process was set to 230 Oe and applied in the longitudinal direction of the sample. Magnetic hysteresis loops were measured using longitudinal magneto-optic Kerr effect (MOKE). For the MI measurements a 4 cm ribbon with preserved width and length was used and an external magnetic field produced by a solenoid applied along the ribbon axis and the impedance was measured by means of the four-point probe method. The ac current passed through the longitudinal direction of the ribbon with different frequencies supplied by a function generator, with a 50 Ω resistor in the circuit. The impedance was evaluated by measuring the voltage and current across the sample using a digital oscilloscope. The MI ratio is defined as $MI\% = \frac{Z(H)-Z(H_{max})}{Z(H_{max})} \times 100$; where $Z$ refers to the impedance as a function of the external field $H$. The $H_{max}$ is the maximum field applied to the samples during the MI measurement. All measurements were carried out at room temperature.

## Results and discussion

### I. MOKE measurements

We first measured MOKE hysteresis loop of FM/IrMn heterostructures, in order to examine the presence of the EB effect and occurrence of the TE in our sample. Figure 1a shows the MOKE hysteresis loops for the field annealed FM ribbon/IrMn sample and for the FM ribbon that does not show any shift. As shown, the hysteresis loop for field annealed FM ribbon/IrMn clearly exhibits the EB behavior as a shift of the loop towards negative values with $|H_{EB}|$=2.4 Oe accompanied with an enhanced $H_c$. The consecutive (n= number of field sweep) MOKE hysteresis loops of a field annealed (280 °C /230 Oe) FM ribbon/IrMn sample are shown in Figure 1b. It is seen that $H_{EB}$ and $H_c$ decrease with increasing the number of loop cycles. Note that the TE is more prominent in the first two consecutive hysteresis loops and in the descending branch because in our measurement the loop starts at positive saturation the descending curve than in the ascending curve of the hysteresis loop.

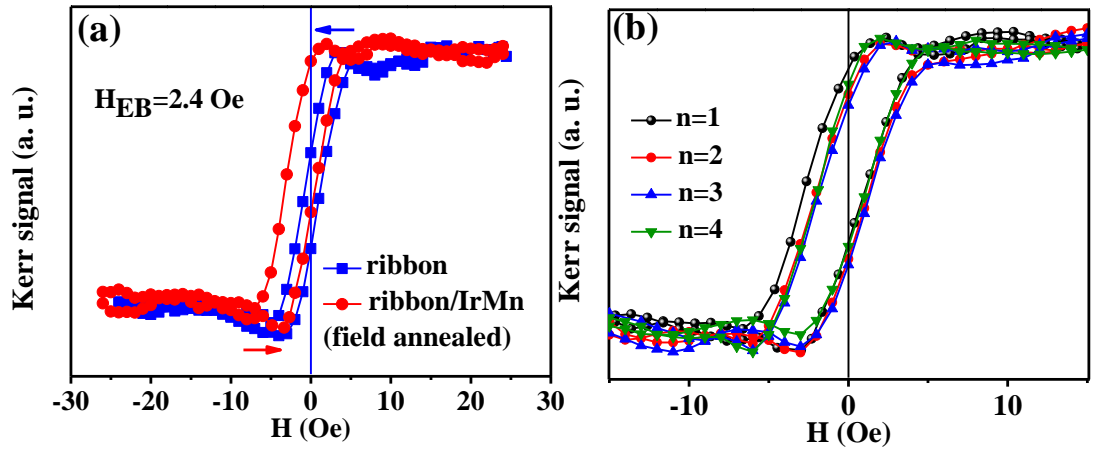

**Figure 1**. (a) MOKE hysteresis loops of the FM ribbon that does not show any shift so that, for field annealed ribbon/IrMn heterostructures, the hysteresis loop clearly exhibits the EB effect by $|H_{EB}|$ =2.4 Oe. (b) Successive MOKE hysteresis loops of the magnetically annealed (280 °C /230 Oe) FM ribbon/IrMn (20 nm) heterostructure at room temperature. Also the hysteresis loop for n=1clearly exhibits the TE between the first two consecutive hysteresis loops.

## II. Effect of EB on MI measurements

The use of EB is particularly attractive in the MI response due to elimination of nonlinear MI response near zero field. For this purpose, there are few reports on the effect of the EB coupling on the MI effect in different structures such as ribbon/oxide layer [38] and also NiFe/FeMn [39] and NiFe/IrMn [40,41] thin film multilayers. Due to presence of the EB in those systems, shifts were observed in both the hysteresis loop and the MI response, and the MI response showed an *asymmetric magnetoimpedance (AMI)*. In section of MOKE measurement, we showed that the both EB and TE is present in our samples. Here, in order to elucidate the effect of the EB coupling on the MI effect, we carry out field sweep impedance measurements and investigate how the impedance of the heterostructures changes as a function of the external magnetic field.

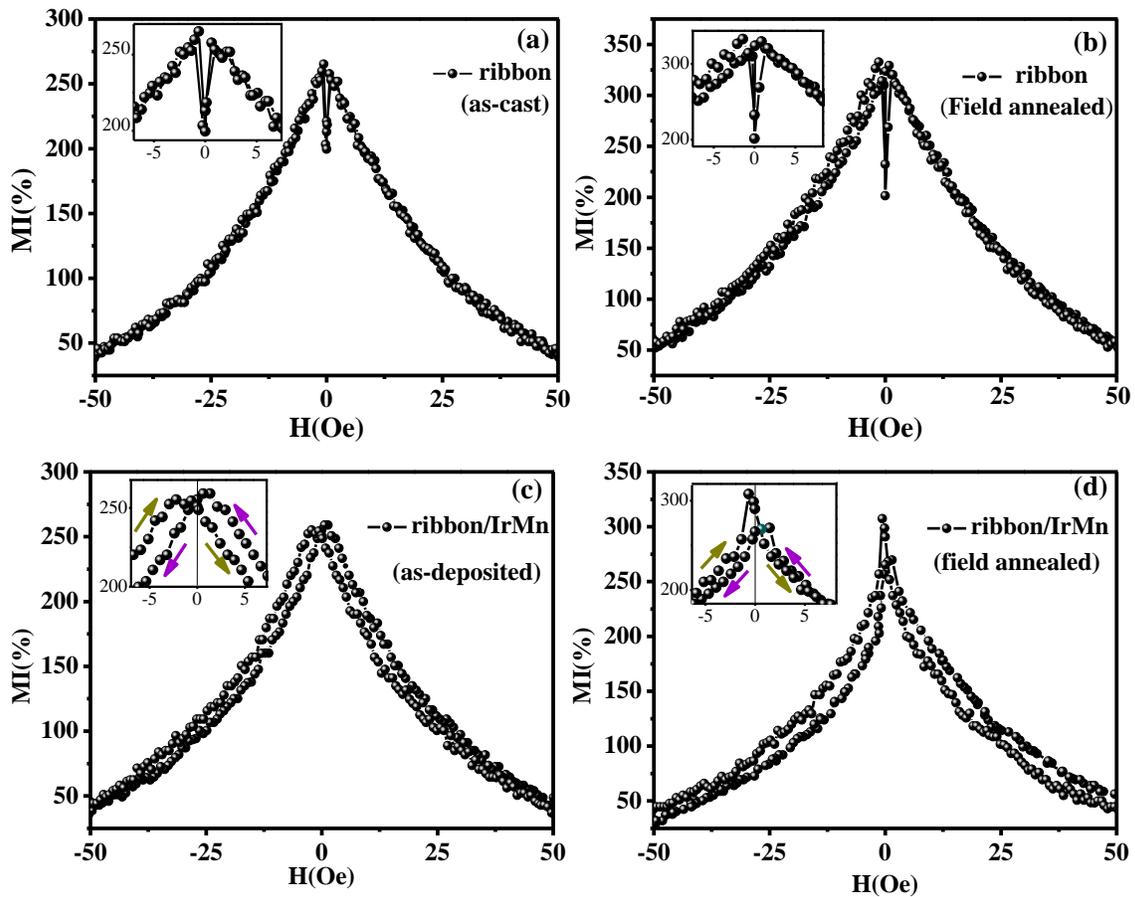

**Figure 2**. Comparison of the MI response of the ribbon and the ribbon/IrMn samples at frequency of $f$ = 10 MHz and an applied current of $I$= 66 mA. Inset shows the MI response around zero magnetic fields.

Figure 2a-b and 2c-d show MI response for ribbon (as-cast and field annealed) and ribbon/IrMn heterostructures (as-deposited and field annealed) samples measured at $f = 10$ MHz and a current of $I = 66$ mA. There are differences in the MI effect measured for the as-cast ribbon and the IrMn-deposited ribbons.

- *i) The decrease of MI response:* MI% is 265% for the as-cast ribbon and decreases to 259% for as-deposited ribbon/IrMn, and MI% is 325% for the field annealed as-cast ribbon and decreases to 314% for the field annealed ribbon/IrMn sample. The decrease of MI response for IrMn-deposited samples is associated with a decrease in the transverse magnetic permeability, $\mu_t$. Decrease in the $\mu_t$ and thereby in MI ratio is due to presence of the EB effect in the IrMn/ribbon interface which pins the spins of the ribbon with additional interface anisotropy. In addition, the as-cast ribbon exhibits a double peak behavior whereas the IrMn deposited ribbon shows a single peak one (Figure 2a and 2c). As can be seen in in Figure 2b and 2d, this behavior is also repeated for field annealed samples. The observed single or double-peak behavior in MI response, generally, is associated with the longitudinal or transverse magnetic anisotropy with respect to the applied external magnetic field direction [42,43]. From these results, we conclude that the transverse magnetic anisotropy of ribbon changes by presence of the IrMn layer.

- *ii) Hysteretic MI behavior:* Contrary to that as-cast ribbons, it can be seen in Figure 2c-d that the IrMn deposited samples show a hysteretic MI behavior. This fact potentially implies presence of an antiferromagnetic coupling at the interface [44]. Also, we have shown in our previous work [29] that the hysteretic behavior of the MI response depends on the exchange coupling strength at the interface. Similarly, in the sample used in this work, exchange coupling at the interface between IrMn and ribbon can result the same effect and so the hysteretic behavior in MI response can be observed.

- *iii) Asymmetry and shift of the MI response:* In the inset of Figure 2d we observe an asymmetric single peak of the MI response with a nonzero $H$ (2.4 Oe) for the ribbons coated with IrMn. This shift is equal to the one observed for magnetization loop as an EB shift ($|H_{EB}|$ =2.4 Oe). The observed shift occurs at all applied frequencies (1-20 MHz, not shown here). As mentioned before, asymmetric MI (AMI) response due to presence of the EB [38] is interpreted as the exchange interaction between the amorphous bulk and the surface crystalline regions and in turn forming an effective unidirectional anisotropy. Here, the exchange interaction between the IrMn and the surface of ribbon can lead to an effective unidirectional anisotropy, as represented in Figure 2d, and is responsible for the AMI.

In addition, in order to study the influence of interface roughness on exchange coupling in our sample, the following discussion are presented. We know that the interface roughness can influence on exchange coupling in FM/AFM, so that produces a random field, which leads to the breakup of the AFM into domains with domain walls perpendicular to the interface and these domains generate uncompensated moments that couple to the FM, resulting in EB [3]. There are no reports on effect of roughness on the EB in ribbon/AFM sample, but there are many reports which show that the hysteresis loop shift can be observed in bulk ribbon materials if they were moderately annealed below the crystallization temperature [38,45–47]. But, the maximum value of EB for ribbon alone when annealed in different temperatures, considering the roughness of the ribbon, was seen to be about 1 A/m = 0.012 Oe with a significant behavior in the MI response in different field polarities [46]. We highlight that our sample (ribbon/IrMn) shows the EB field of 2.4 Oe (from MOKE measurements) that changes the MI behavior and definitely this amount arises from IrMn layer and not caused by the surface roughness. In addition, because the sample was etched before deposition, the interface should be clean, but might be rough. Due to low magnitude of EB field of ribbon alone (0.012 Oe) compared with ribbon/IrMn (2.4 Oe) we conclude that there would not be any variation in the proposed scenario as the roughness is not a variable in the experiments. It is noticeable that the 0.012 Oe field shift occurs by annealing in air (oxidic crystallization of surface ribbon) whereas, in our sample annealing carried out in vacuum condition. Therefore, oxidation cannot occur in our sample and the exchange is just due to presence of IrMn layer.

**III. Modeling and consistency of the MI and MOKE measurements**

We speculate that such a coupling mechanism with the observed AMI response is similar to those of soft/hard bilayer exchange spring magnets [48]. The MI effect is mediated by the ac current skin depth with a thickness of approximately 100 nm. The interface of FM ribbon in the skin depth region affected by the EB has a different FM property to the rest of it. A region close to the interface pins stronger than it's underneath and behaves as an exchange spring medium. Schematic of spin configuration of ribbon/IrMn heterostructures at the interface with exchange coupling can be seen in Figure 3. This Figure shows the evolution of spin configuration during external field ($H$) reversal from the opposite of EB induced field ($H_{EB}$) to its parallel orientation, in the skin depth ($\delta$) region. Each step in Figure 3 is shown as panel numbers, referring to a corresponding reversal point addressed in Figure 3b, c that show MI response and first (n= 1) MOKE hysteresis loop of field annealed ribbon/IrMn, respectively. Step (1): the external field H has a maximum value and thereby all spins in FM layer are parallel to the external field H. Step (2): By decreasing the external field H from positive saturation, from right to left, according to the exchange interaction between the spins of the IrMn layer and ribbon and presence of $H_{EB}$ in opposite direction of external magnetic field, highest value of $\mu_t$ occurs in this step and MI has its peak at this point and no peak at H = 0 Oe. A ferromagnetic exchange spring happens in this step. Step (3): When H=0, according to the MOKE measurement, switching of the spins does not happen completely. Step (4): When H is reversed the spins of FM ribbon completely switch. Therefore, we have exchange spring in FM layer and it is consistent with the MI and MOKE measured data.

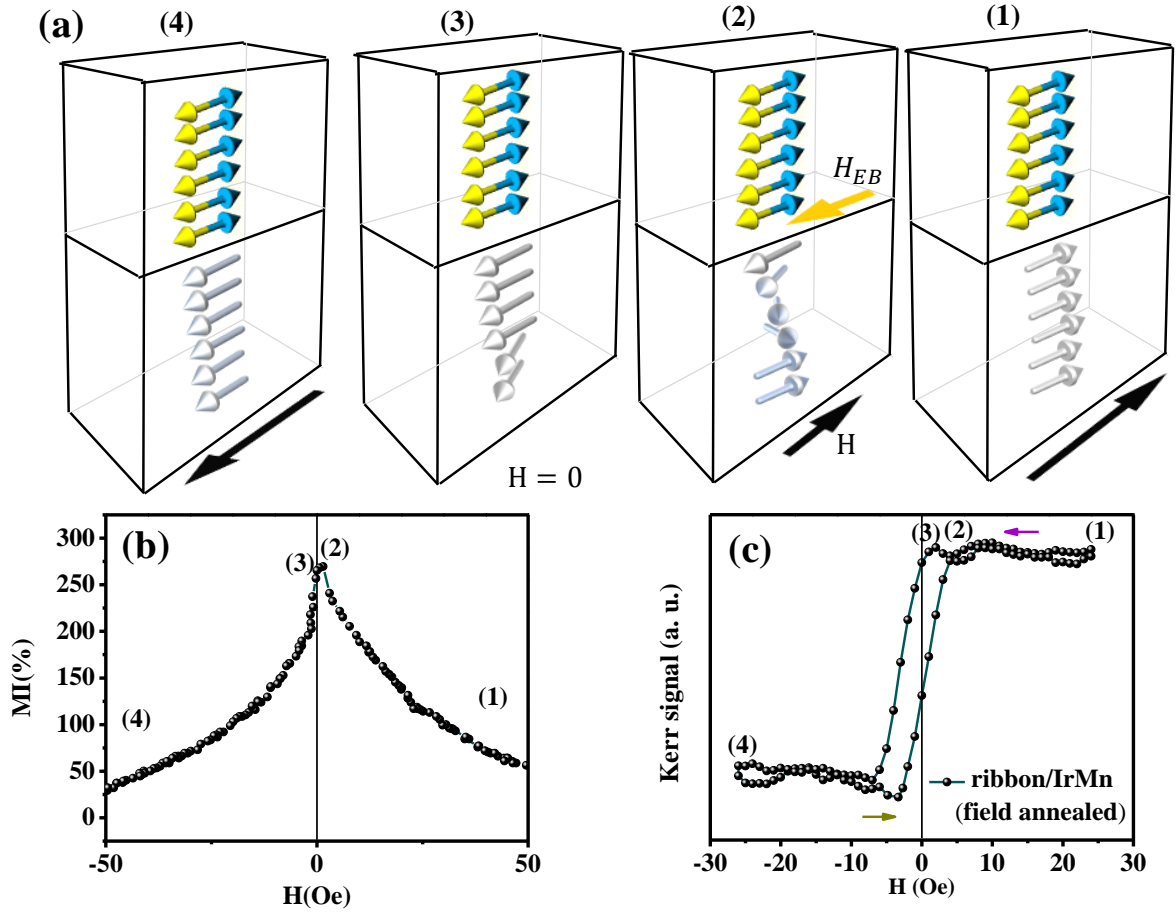

**Figure 3**. (a) Schematic of the evolution of spin configuration during external magnetic field ($H$) reversal from the opposite of exchange bias ($H_{EB}$) to its parallel direction in FM/AFM heterostructures in the skin depth ($\delta$) region. Length of the black arrows shows the magnetic field strength. (b) MI response of field annealed ribbon/IrMn. (c) First MOKE hysteresis loop of the magnetically annealed (280 °C /230 Oe) FM ribbon/IrMn (20 nm) heterostructure at room temperature

## IV. TE probed by MI

In order to investigate of TE through the MI effect, we carry out the successive field cycling while performing MI measurements for the field annealed ribbon/IrMn samples. For mentioned samples consecutive MI curves are presented in Figure 4a-d. Measurement is done at frequency of $f$ = 10 MHz and labeled by n = 1 to n = 4, respectively from first to fourth sequential MI measurements. By continuously repeating the MI measurements, the peaks which appeared at positive and negative low magnetic fields move towards $H$= 0 Oe. Also, the asymmetric behavior of the MI response disappears. This result obviously origins from the TE at the interface of the ribbon/IrMn. The strength of exchange interaction decreases due to TE and finally MI effect shows symmetric behavior after continuous MI measurements. No additional changes appear in the MI plot by further repeating the measurement.

We should also that the MI% increases as well by repeating the measurement from the first to fourth cycle. As mentioned before, the MI% is dominated by the strength of $\mu_t$. In fact, the $\mu_t$ of the ribbon increases due to unpinning of the magnetic moments at TE and consequently the MI% increases. Therefore, peak shift and 52% MI% increase observed due to TE. We repeat the experiments for a field annealed ribbon without IrMn and no variation evidenced in MI response by repetitive measurements.

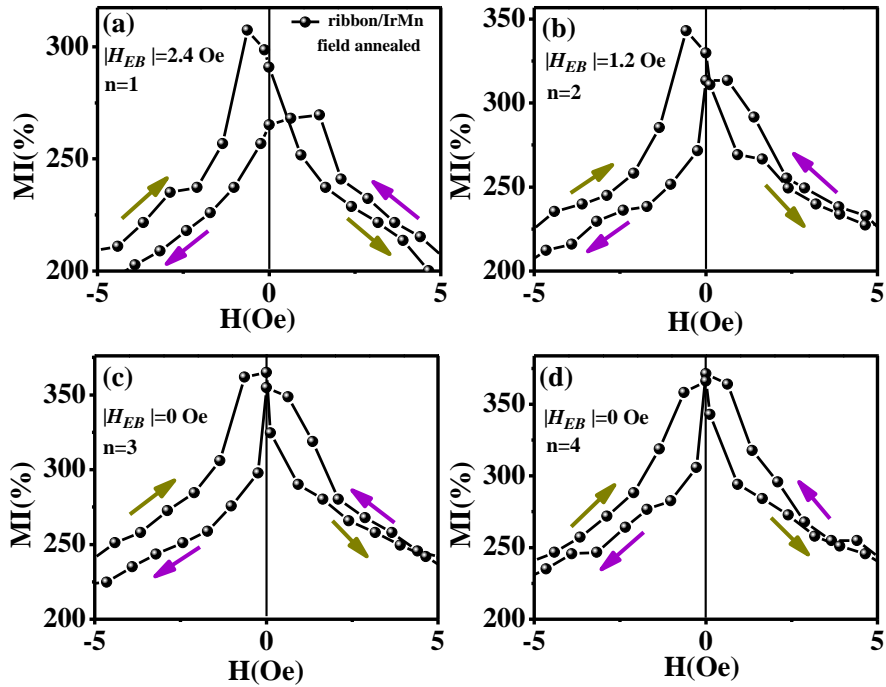

**Figure 4**. MI response of field annealed ribbon/IrMn for 4 consecutive MI loops at $f$ =10 MHz. By repeating the MI measurements continuously (n = 1 to n = 4), the peak which is appearing at low magnetic fields moves towards the $H$= 0 Oe and finally the MI response at n=4 shows increased value with a symmetric and single peak behavior.

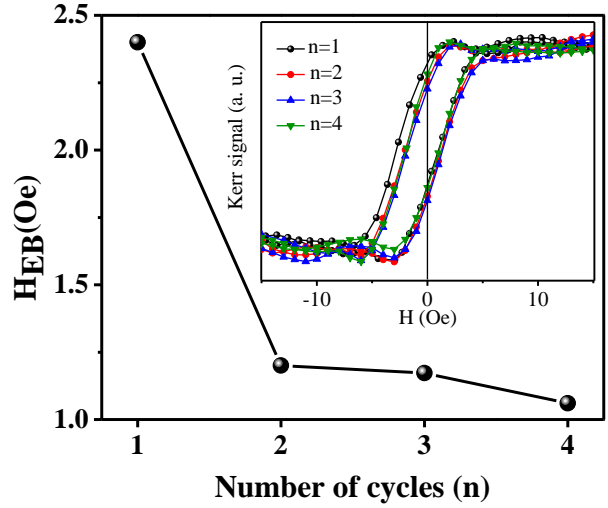

**Figure 5**. Variation of $H_{EB}$ as a function of number of cycles (n) obtained from MOKE hysteresis loops.

Generally, in EB systems two different TE, viz., athermal and thermal mechanisms, has been considered as in order to separate the distinct feature of the first hysteresis cycle [13,16,17,49]. Athermal TE is reported particularly in EB systems with high symmetry antiferromagnets in their structure, such as, NiMn, FeMn, and IrMn [10]. Therefore, as we have used IrMn here, we can analyze the EB training effect in terms of athermal training which includes spin-flop like coupling because of the AFM biaxial anisotropy in the system [10]. A rapid decrease in $H_{EB}$ between n=1 and n=2 can be seen in Figure 5. As previously modeled by Hoffmann [10], an abrupt decrease in the EB between the first and second cycles shows that TE has athermal contribution. In fact, Hoffman predicted that athermal TE is due to the initial stabilization of noncollinear arrangement of AFM sublattice spins after field cooling via a spin-flop like coupling mechanism at the interface. After first field cycling the spins of the AFM rearrangement into a collinear style and gives rise to an abrupt decrease in the $H_{EB}$. We believe that Hoffmann's model offers a good description of our experimental data, due to the existence of a strong single cycle TE in our IrMn/FM system.

**Conclusion**

In summary, we probed the exchange bias training effect through the MI measurement in IrMn-layer/ferromagnetic-ribbon heterostructures. The AFM layer creates a bias field in the ferromagnetic-ribbon that shifts the hysteresis loop and the MI response and also creates asymmetric and hysteretic behavior in MI response. It is observed that, during the training procedure in field sweep MI measurement, both EB field and the degree of asymmetry of MI decrease. In addition, we proposed occurring exchange spring model in FM according to the MI and MOKE measurements. The significant decrease in $H_{EB}$ taking place between n=1 and n=2 field sweep, can be microscopically explained within the framework of Hoffmann's model.

**Acknowledgments**

Support from Iran National Science Foundation (INSF) is acknowledged.